\documentclass[useAMS,usenatbib,referee,usegraphicx]{mn2e}

\title[Characteristics of supernova remnant G351.7+0.8]{Characteristics of supernova remnant G351.7+0.8 and a distance argument against its association with PSR J1721-3532}
\author[W. W. Tian, M. Haverkorn and H. Y. Zhang]{W. W. Tian$^{1,2}$\thanks{E-mail:
tww@iras.ucalgary.ca; tww@bao.ac.cn}, M. Haverkorn$^{3,4}$, H. Y. Zhang$^{1}$\\
$^{1}$National Astronomical Observatories, CAS, Beijing 100012, China\\
$^{2}$Department of Physics $\&$ Astronomy, University of Calgary, Calgary, 
Alberta T2N 1N4, Canada\\
$^{3}$Jansky Fellow, National Radio Astronomy Observatory; marijke@astro.berkeley.edu \\
$^{4}$Astronomy Department, University of California-Berkeley, 601 Campbell Hall, Berkeley CA 94720, USA}
\begin{document}

\date{Accepted . Received ; in original form }

\pagerange{\pageref{firstpage}--\pageref{lastpage}} \pubyear{}

\maketitle

\label{firstpage}

\begin{abstract}

New images of the supernova remnant (SNR) G351.7+0.8 are presented
based on 21cm HI-line emission and continuum emission data from the
Southern Galactic Plane Survey (SGPS).  SNR G351.7+0.8 has a flux
density of 8.4$\pm$0.7~Jy at 1420~MHz. Its spectral index is
0.52$\pm$0.25 (S=$v$$^{-\alpha}$) between 1420 MHz and 843 MHz, typical of adiabatically
expanding shell-like remnants. HI observations show structures
possibly associated with the SNR in the radial velocity range of $-10$
to $-18$~km$/$s, and suggest a distance of 13.2~kpc and a radius of
30.7~pc. The estimated Sedov age for G351.7+0.8 is less than
6.8~$\times$10$^{4}$ yrs.  A young radio pulsar PSR J1721-3532 lies
close to SNR G351.7+0.8 on the sky.  The new distance and age of
G351.7+0.8 and recent proper-motion measurements of the pulsar
strongly argue against an association between SNR G351.7+0.8 and PSR
J1721-3532. There is an unidentified, faint X-ray point source 1RXS
J172055.3-353937 which is close to G351.7+0.8. This may be a neutron
star potentially associated with G351.7+0.8.
\end{abstract}


\section{Introduction}
Supernova remnant (SNR) and pulsar (PSR) associations have been widely
studied since the first radio pulsar was discovered in
1968. Associations are usually judged on criteria such as agreement in
age and distance, angular proximity, and whether the transverse
velocity derived for the pulsar is reasonable. However, an independent
measure of the age and distance for the two objects is usually
difficult. Often, the error in the measurement of the transverse
velocity of a pulsar is comparable to the measurement itself,
effectively rendering a non-detection.  Another
recently much-used method of confirming an association is searching
for a pulsar wind nebula (PWN) in a SNR. If a pulsar is still inside
its parent SNR, the high pressure of the surrounding plasma can
confine the pulsar's relativistic wind generating synchrotron emission
observable as a PWN.  About 50 possible associations of
SNR/PSR have been claimed in the literature (for a compilation see
Tian $\&$ Leahy 2004, and additionally Tian $\&$ Leahy 2006), and a
few of these associations have been contested (Nicastro et al. 1996;
Stappers et al. 1999; Gaensler et al. 2001; Thorsett et al. 2002;
Brisken et al. 2006).

G351.7+0.8 is a shell-type SNR first mapped by Whiteoak $\&$
Green (1996). It is a faint plateau of emission on the edge of a
bright HII region.  Gaensler $\&$ Johnston (1995) suggested a possible
association with pulsar PSR J1721-3532, which was included in the
SNR/PSR associations catalogue by Tian $\&$ Leahy (2004).  In this
paper, based on new radio observations of SNR G351.7+0.8 and recent
proper-motion measurements of pulsar PSR J1721-3532, we give new
physical parameters of G351.7+0.8 and argue against the association
SNR G351.7+0.8/PSR J1721-3532.

\section{Observations}
The radio continuum and HI emission data sets come from the Southern
Galactic Plane Survey (SGPS) which is described in detail by Haverkorn
et al. (2006) and McClure-Griffiths et al. (2005).  The SGPS images
the HI line emission and 1.4~GHz continuum in the range 253$^{\circ} <
l < 357^{\circ}$, mostly in the fourth quadrant of our Galaxy,  with the
Australia Telescope Compact Array (ATCA) and the Parkes 64m single
dish telescope. The continuum observations have a resolution of
100~arcsec and a sensitivity better than 1~mJy/beam. The HI data have
an angular resolution of 2~arcmin, a rms sensitivity of $\sim$1~K and
the velocity resolution of the SGPS of 1~km~s$^{-1}$.  The continuum
data are taken with the ATCA interferometer only, which is insensitive
to scales larger than 30~arcmin. Because the angular size of the SNR
(18~arcmin$\times14$~arcmin in l$\times$b) is significantly smaller
than this, all remnant emission is included in the continuum data.
                                                                                
\section{Results}
\subsection{Continuum Emission}
\begin{figure}
\vspace{50mm}
\begin{picture}(0,50)
\put(-85,245){\includegraphics{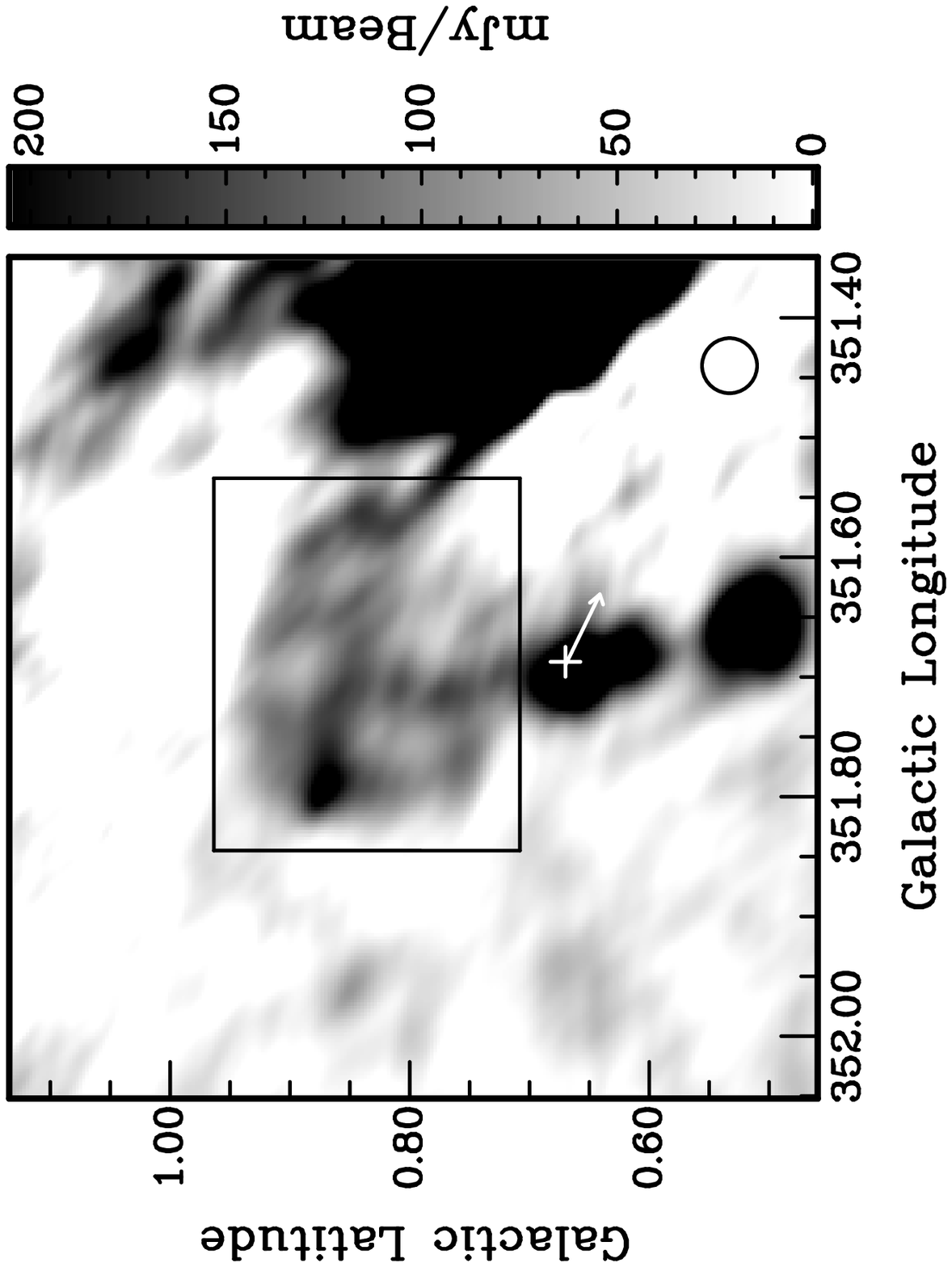}}
\end{picture}
\caption[xx]{The images of G351.7+0.8 at 1420 MHz from the SGPS with a
beamsize of 100~arcsec, as shown in the lower right corner of the
plot.  The white cross and connected arrow show the position and
proper motion direction of the pulsar PSR J1721-3532. The box in the
map is used to estimate the flux density of G351.7+0.8.}
\end{figure}

Fig. 1 shows the image of G351.7+0.8 at 1420 MHz from the SGPS. The
SGPS image has a lower resolution (100 arcsec) but higher sensitivity
($<$1~mJy/beam) than a previously published 843~MHz map (43~arcsec and
$>$2~mJy/beam) from the Molonglo Observatory Synthesis Telescope
(MOST) (Whiteoak $\&$ Green 1996), so the SGPS image shows more faint
emission than the MOST map. The 1420~MHz and 843~MHz images have
similar morphologies. The flux density of G351.7+0.8 is 8.4$\pm$0.7~Jy
at 1420~MHz and 11$\pm$1.1~Jy at 843~MHz (Whiteoak $\&$ Green 1996),
so that the spectral index between these two frequencies is
0.52$\pm$0.25 (S=$v$$^{-\alpha}$).
The Parkes 6 cm  southern Galactic plane survey by Haynes,
Caswell $\&$ Simons (1978) has such a low resolution (4.1$^{\prime}$)
that a flux estimate is very uncertain due to contamination from the
neighboring HII region, but a rough estimate of the flux from
G351.7+0.8 ($\sim$~2~Jy) agrees well with the estimated spectral
index.

\subsection{Polarization}
G351.7+0.8 does not show any polarization at 1.4~GHz. As the
non-thermal emission from supernova remnants is intrinsically
partially polarized, the lack of polarization indicates that the
emission is depolarized. This depolarization is most likely due to
varying polarization angle on scales smaller than the synthesized
beam, caused by small-scale structure in electron density and/or
magnetic field in or in front of the SNR. As SNRs at larger distances
tend to exhibit structure on smaller angular scales, SNRs beyond a
certain distance appear mostly depolarized.  Comparison with many
remnants with known dispersion measure (DM) indicates that G351.7+0.8
(DM$=979$~pc~cm$^{-3}$ for a distance of 13.2~kpc, see section 4.2),
is expected to be completely depolarized (Kothes \& Landecker 2004), as observed.
                           
\subsection{HI Emission}
                                                                        
\begin{figure*}
\vspace{150mm}
\begin{picture}(60,80)
\put(-235,540){\includegraphics{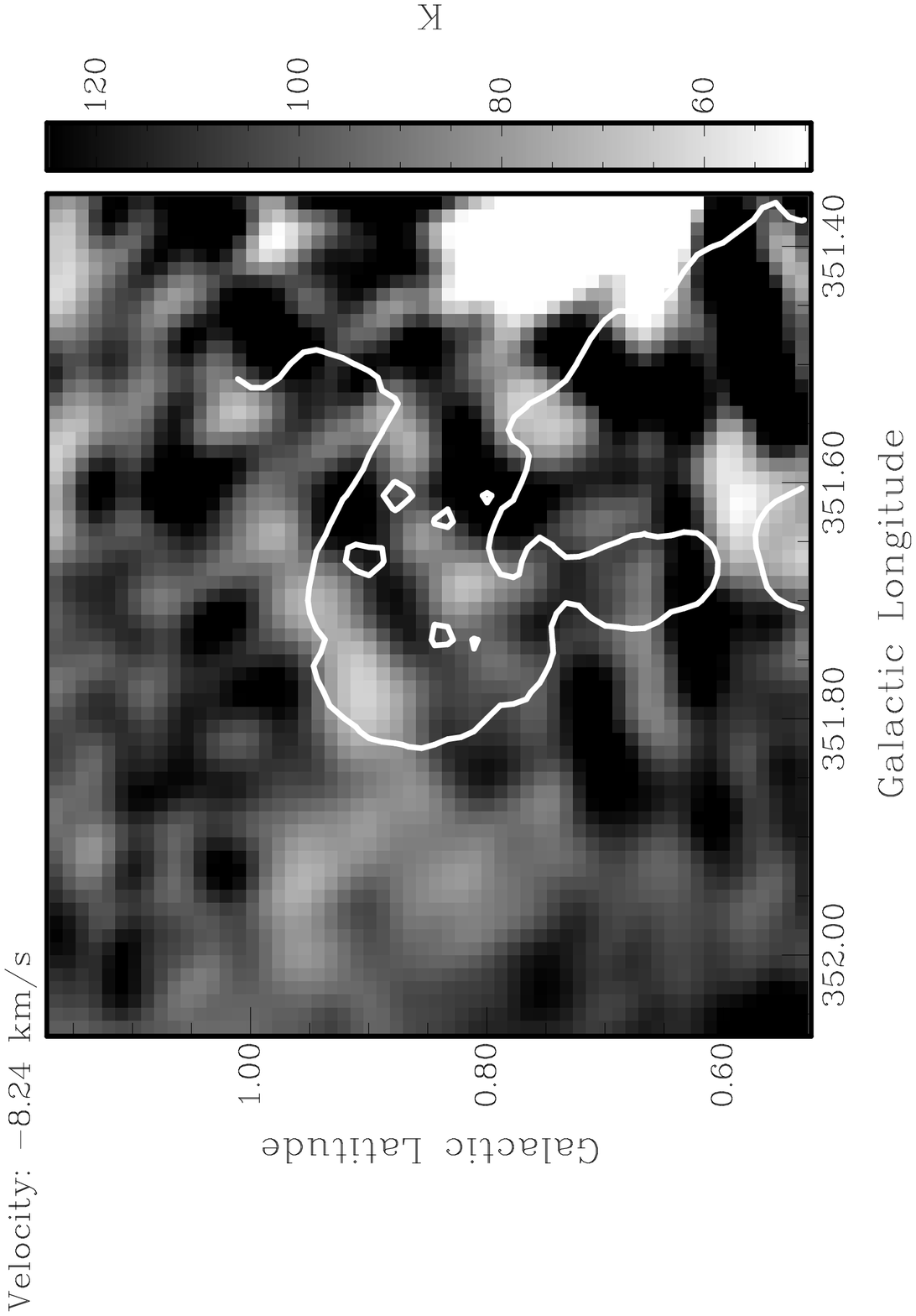}}
\put(50,520){\includegraphics{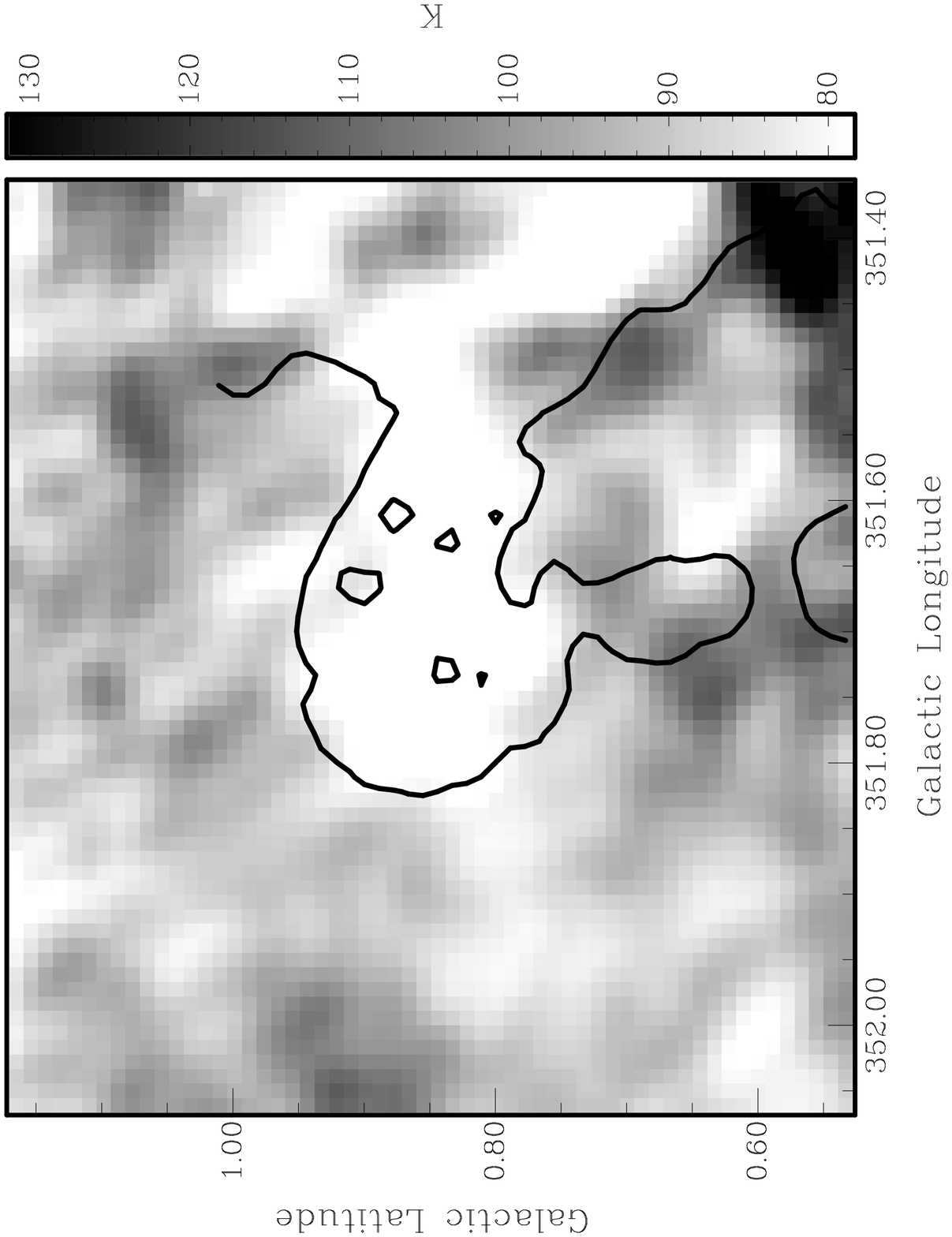}}
\put(-210,345){\includegraphics{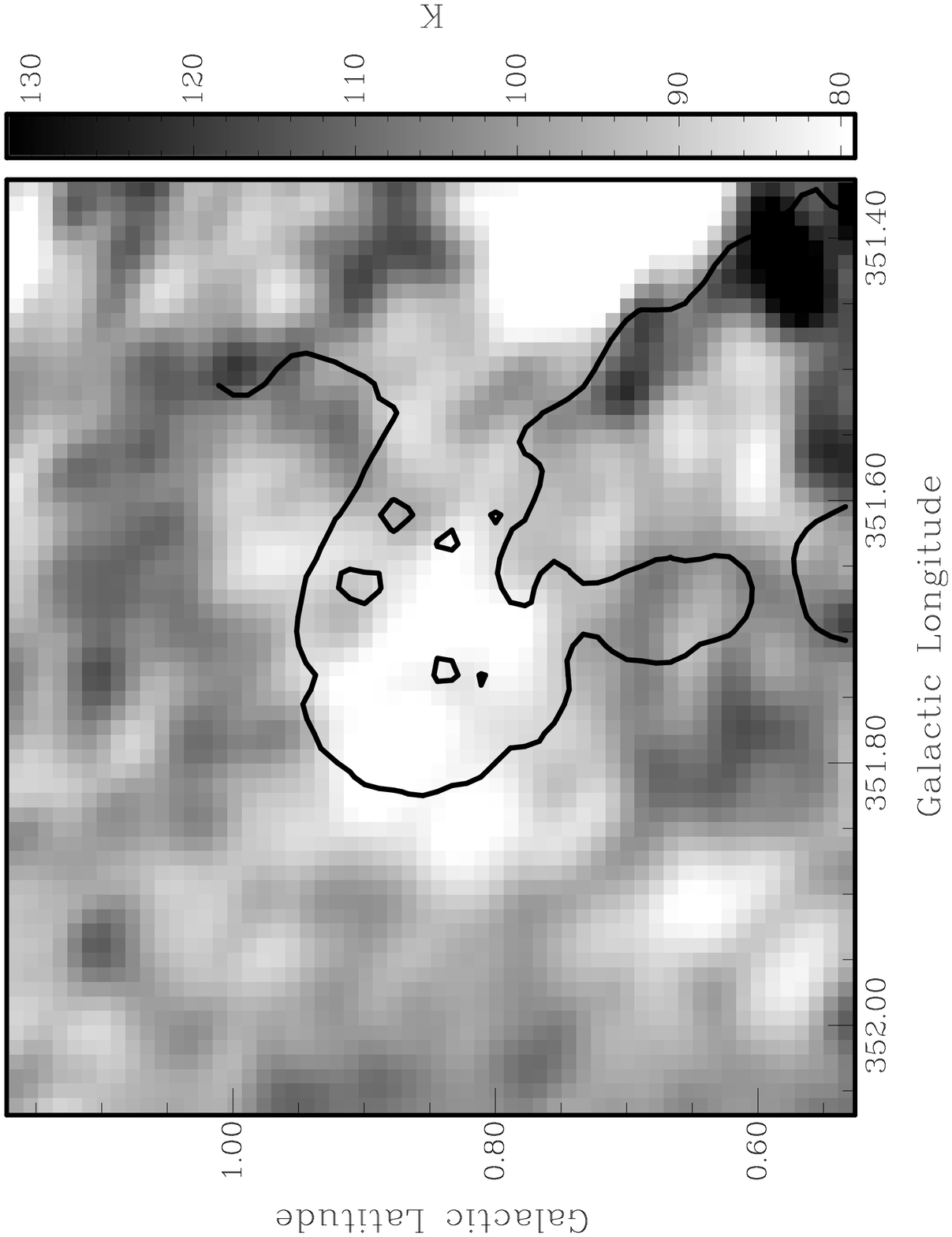}}
\put(50,345){\includegraphics{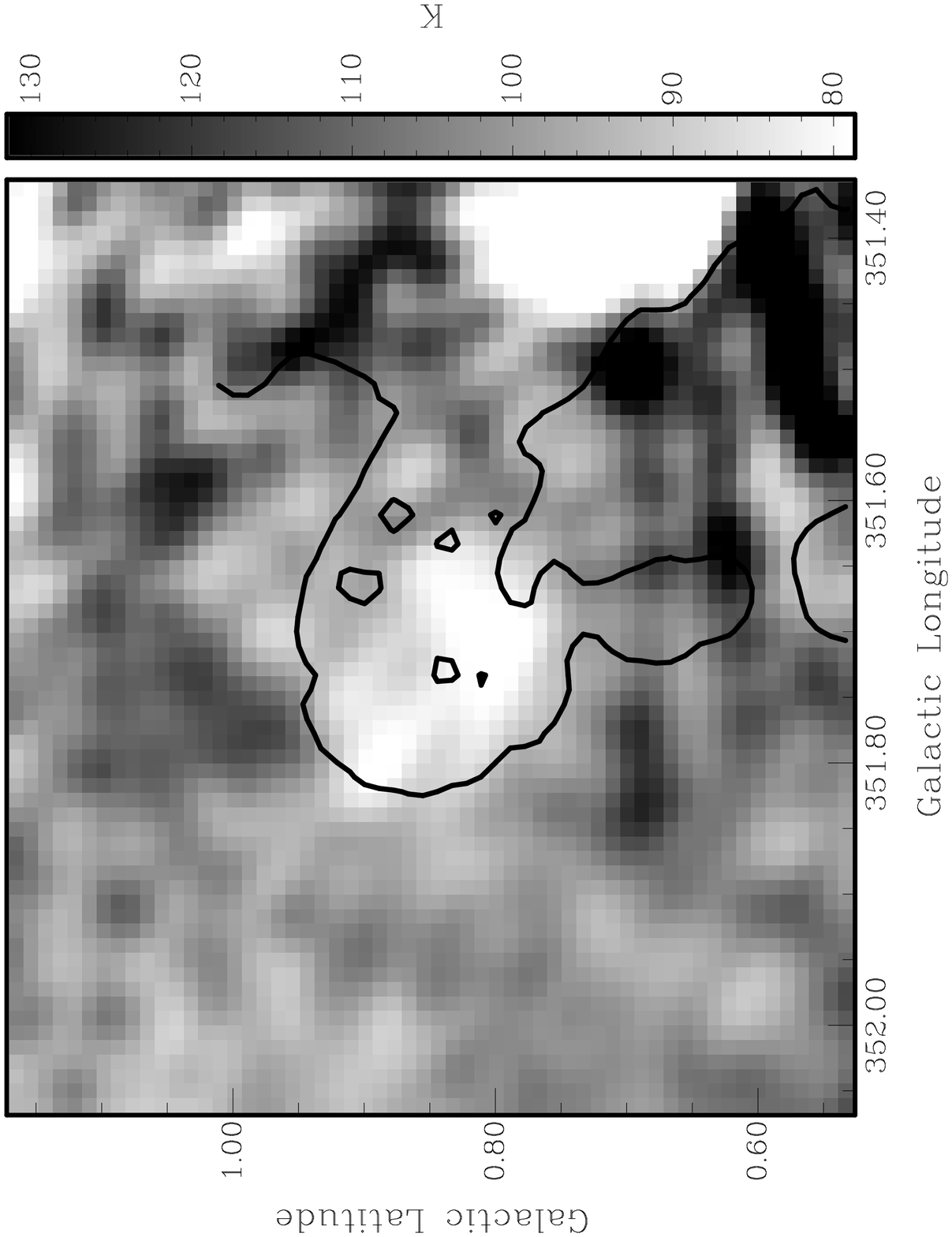}}
\put(-245,195){\includegraphics{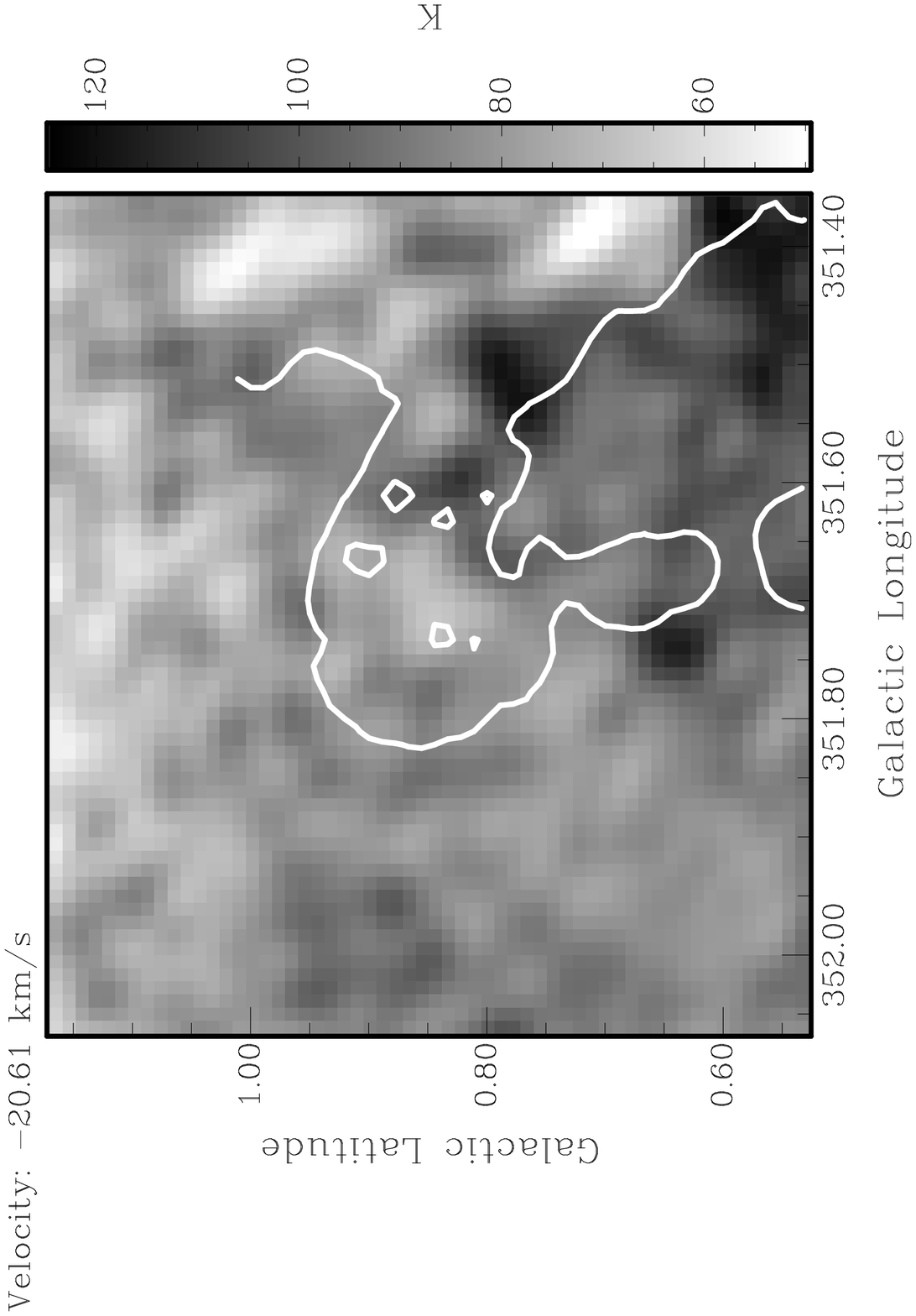}}
\put(65, 160){\includegraphics{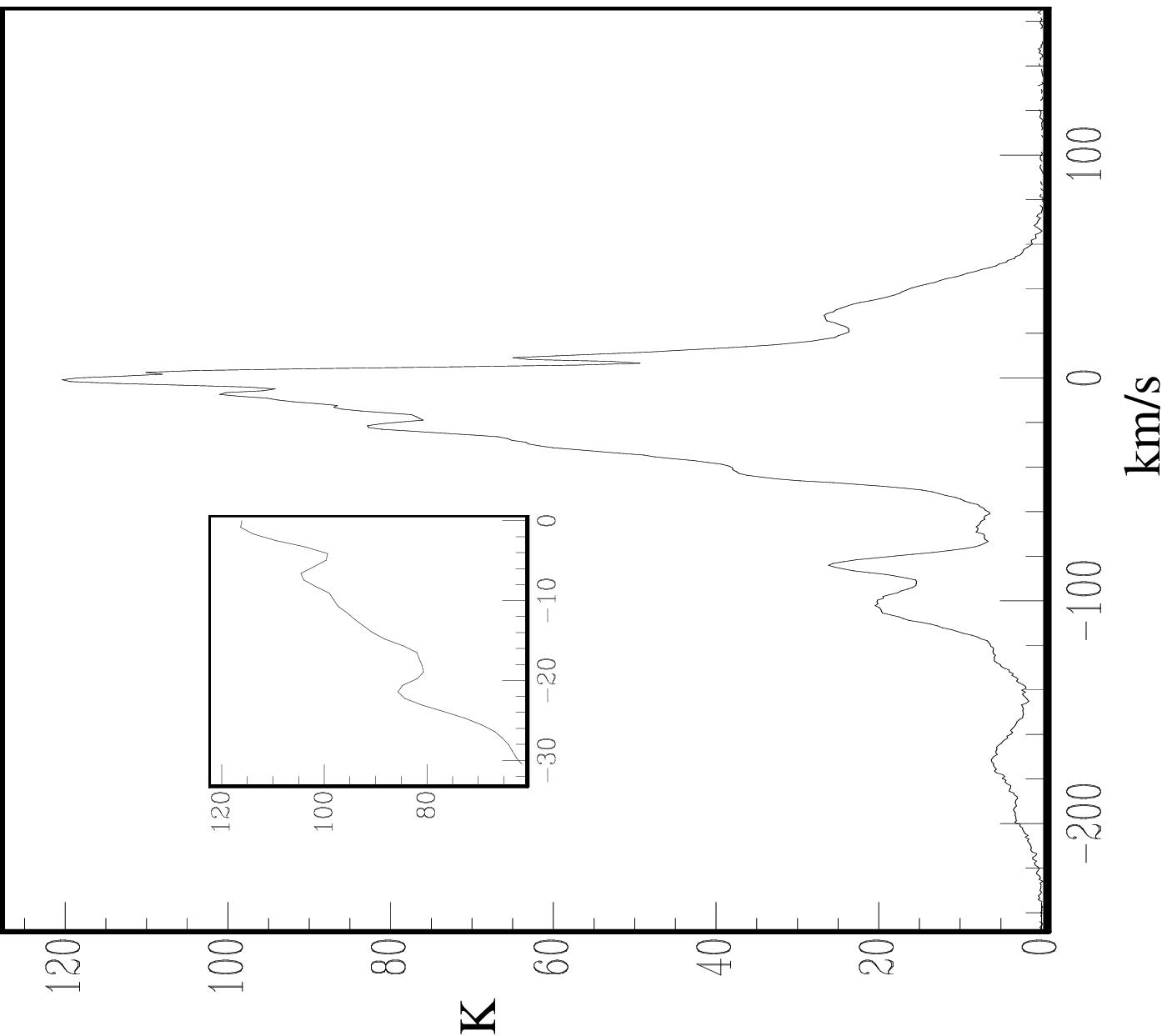}}

\end{picture}
\caption[xx]{HI emission in the field centered on G351.7+0.8. The
    first five frames (left to right, top to bottom) show HI emission
    at -8.24~km~s$^{-1}$, averaged over velocity ranges [-11.54,
    -13.19], [-14.01,-15.66], and [-16.49,-18.14]~km~s$^{-1}$, and at
    -20.61~km~s$^{-1}$, respectively.
A contour of the 1420~MHz continuum at 58~mJy/beam overlaid on each
map. The bottom right plot shows the HI spectrum over the complete
velocity range integrated over the whole remnant. The inlay shows the
velocity region in which the HI is morphologically similar to the
continuum emission from the remnant.}
\end{figure*}

\begin{figure}
\vspace{50mm}
\begin{picture}(0,50)
\put(-25,200){\includegraphics{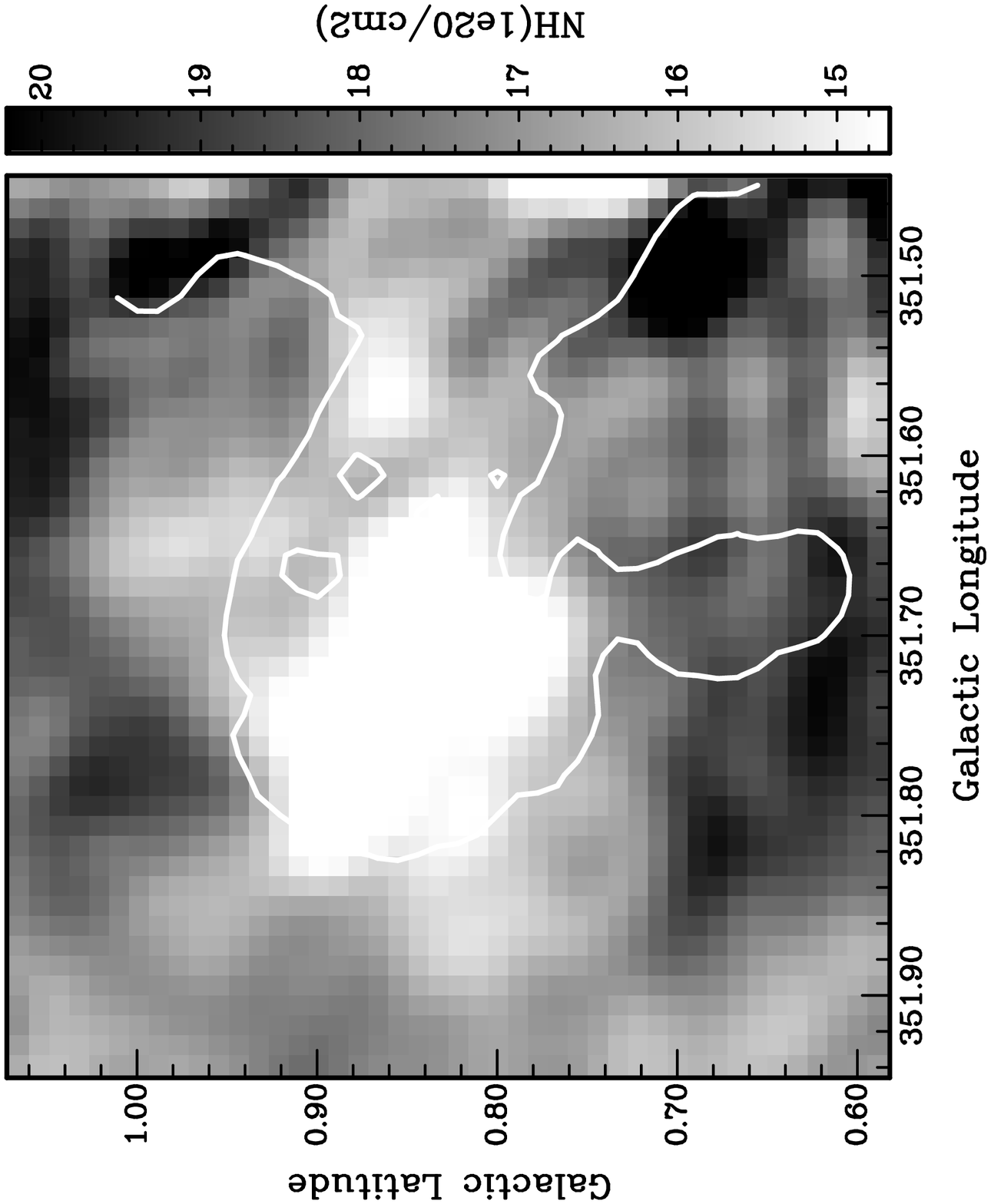}}
\end{picture}
\caption[xx]{Column density of the HI in the field centered on
G351.7+0.8, in 10$^{20}$ atoms~cm$^{-2}$. The contour is same as
in Fig.~2.}
\end{figure}
                                                                               
We searched the SGPS radial velocity range from -247 km~s$^{-1}$ to
165 km~s$^{-1}$ for features in HI with similar morphology as
G351.7+0.8. There is emission which is possibly coincident with the
SNR only in the velocity range -10 to -18 km~s$^{-1}$.  Fig.~2 shows
maps of HI emission associated with the SNR averaged over velocities
in the ranges of [-11.54, -13.19], [-14.01,-15.66], and
[-16.49,-18.14]~km~s$^{-1}$, flanked by two off-line channels. The HI
spectrum over the whole velocity range, integrated over the whole
remnant, is also shown. Superimposed on each map is a contour of 1420~MHz 
continuum at 58
mJy/beam , chosen to denote the position and
extent of the SNR.  Fig. 3 shows the column density map of the HI
integrated over channels from -10 to -18 km/s. Continuum subtraction
has been performed in all plots.

\section{Discussion}
\subsection{The Distance and age}
Using a flat Galactic rotation velocity V$_{R}$=V$_{0}$=200
km~s$^{-1}$ and R$_{0}$=8.0~kpc, we obtain the SNR's distance
$d=2.7\pm$0.6~kpc or $d=13.2\pm$0.5~kpc.  The angular size of
G351.7+0.8 is 18$^{\prime}$ by 14$^{\prime}$ (average diameter
16$^{\prime}$), so this yields a radius of about 6 pc ($d=2.7$~kpc) or
30.7~pc ($d=13.2$~kpc) for the SNR.  From the column density of the HI
associated with G351.7+0.8 (see Fig. 3), we estimate an approximate
N$_{H}$ of $1-5\times$$10^{20}$~cm$^{-2}$, so the approximate density
n$_{0}$=N$_{H}$/(2R) is about 7.8~cm$^{-3}$ ($d=2.7$~kpc) or 1.6
cm$^{-3}$ ($d=13.2$~kpc).  An ISM density of 7.8~cm$^{-3}$ is so high
that we should have seen HI piled up in a clear shell around the SNR,
which is not seen in Fig. 2. The absence of such a shell means
G351.7+0.8 is most likely at the farther distance of 13.2~kpc.

Applying a Sedov model (Cox 1972), for a typical explosion energy of
$E=0.5\times$$10^{51}$~erg ($\epsilon_{0}=E/(0.75\times
10^{51}$~erg)=2/3), yields an age of 1.4$\times$10$^{5}$~yr for
$d=13.2$~kpc.

Because the Sedov phase only lasts for several 10,000 years, the SNR
would have completely cooled by that time for n$_0$=1.6~cm$^{-3}$.
The shock radius when the SNR has cooled is
$R_s^{(c)}$=24.3($\epsilon_{0}/n_{0}$)$^{5/17}n_{0}^{-2/17}$~pc.
Because the remnant would show a thin, bright shell if it was older
than a cooling time, contrary to observations, we require that the SNR
age is smaller than the cooling time, i.e.\ R$\le R_s^{(c)}$. Thus we
find an upper limit for n$_{0}$ of 0.4 cm$^{-3}$.  This is consistent
with a HI column density of about 0.8$\times$$10^{20}cm^{-2}$,
roughly consistent with Fig.~3.  The revised age for G351.7+0.8
with n$_{0}$$<$0.4 cm$^{-3}$ is less than 6.8$\times$10$^{4}$ yr.

\subsection{SNR G351.7+0.8 and PSR J1721-3532}

Pulsar J1721-3532 (l=351.69$^{0}$, b=0.67$^{0}$) is situated at a
slightly lower latitude than SNR G351.7+0.8 and obscured by a small
HII region. It has a characteristic age of 178~kyr (Taylor et al. 1993)
and a DM distance of 5.6 kpc based on the electron density model of
Cordes \& Lazio (2002). The pulsar is shown by the white cross in
Fig.~1, where the arrow denotes its proper motion direction. The
angular separation between the pulsar and the centre of the SNR is
$\sim$10$^{\prime}$. Weisberg et al. (1995) gave a lower limit to the
distance of the pulsar of 4.8$\pm$0.3 kpc from HI absorption
spectra. Although no distance and age are available for SNR
G351.7+0.8, Gaensler and Johnston (1995) mentioned a possible
association of SNR G351.7+0.8 with PSR J1721-3532 based on their
$\beta$ parameter method, i.e., their calculations showed a pulsar
this old could have traveled outside the SNR in its lifetime. However,
the SNR's parameters given in the paper make this association
unlikely.
                                                                               
The Cordes $\&$ Lazio (2002) electron density model yields
DM~$=979$~pc~cm$^{-3}$ for the SNR at a distance of 13.2~kpc. The
estimated distance error in the DM values is about 5~per~cent in this
direction (Cordes $\&$ Lazio 2002). The pulsar J1721-3532 has a DM of
496$\pm4$~pc~cm$^{-3}$, which is widely inconsistent with the DM
estimate for the SNR.

\subsection{Timing Measurements for PSR J1721-3532}
A pulsar proper motion measurement is a good way to support SNR/PSR
associations or disprove candidate associations. Recent timing
measurements (Zou et al. 2005) gave PSR J1721-3532's proper motion
7200$\pm$2400 km~s$^{-1}$ at a distance of 5.6~kpc
($\mu_{\alpha}$=-43$\pm$24 $~mas$~yr$^{-1}$,
$\mu_{\delta}$=-270$\pm$93 $~mas$~yr$^{-1}$). Zou et al.'s (2005)
proper motion measurements are based upon comparing two position
estimates (probably including a significant underestimate of position
uncertainties) and cite large errors which are probably related with a
lot of timing noise from the pulsar, adding to the position
uncertainty. However, the
proper motion direction is still more or less similar if the true values
of the velocity are within 2$\sigma$ from the cited value, which adds weak
evidence against the association PSR J1721-3532/SNR G351.7+0.8 (see Fig. 1).

\subsection{The Rosat point source 1RXS J172055.3-353937}

G351.7+0.8 is an old SNR, its soft X-ray emission is not expected to
be bright in a lower density environment, and should have been
completely absorbed at a distance of 13.2~kpc. This is consistent with
lack of detection within G351.7+0.8 in the Rosat All-sky Survey.
There is an unidentified, faint X-ray point source, listed by Voges et
al. (1999) as 1RXS J172055.3-353937. It is centred on
$l$=351.52$^{0}$, $b$=0.71$^{0}$, and lies close to the southeastern
edge of G351.7+0.8. This X-ray source has a total of 6 counts in the  Rosat
All-Sky survey, not enough to deduce any spectral information.  The
source does not seem to have an optical counterpart. It may coincide
with a hard X-ray source in Fig.~2 of Ezoe et al (2006), possibly
indicating a large distance at which soft X-rays have been
absorbed. Obviously, deeper X-ray observations are needed. If it is
confirmed to be a neutron star later, it is reasonable to associate
G351.7+0.8 with 1RXS J172055.3-353937.

\section{Conclusion}
We present new images of the SNR G351.7+0.8 from the 1.4~GHz radio
continuum and HI emission dataset from the SGPS. Its continuum flux
density at 1.4~GHz is 8.4$\pm$0.7 Jy and its spectral index is
0.52$\pm$0.25. Based on associated HI emission, the remnant's distance
is estimated to be $13.2\pm0.5$~kpc. No polarization is detected,
consistent with the expectation of complete depolarization at this
distance towards the Galactic centre. Based on the assumption that the
remnant must be younger than a cooling time, its age is younger than
$6.8\times10^{4}$~yrs, and the ambient ISM density is $n_0 <
0.4$~cm$^{-3}$.
Discrepancies in distance and dispersion measure, and a tentative
proper motion measurement of pulsar J1721-3532, make an association
between the two objects SNR G351.7+0.8 and PSR J1721-3532 unlikely. A
faint unidentified X-ray point source, 1RXS J172055.3-353937, is a
potential candidate to be associated with G351.7+0.8.
                                                                                
\section*{Acknowledgments}
We thank the referee for his/her comments that have improved the paper.
TWW acknowledges supports from the Natural Sciences Foundation of
China and the Natural Sciences and Engineering Research Council of
Canada. MH acknowledges support from the National Radio Astronomy
Observatory (NRAO), which is operated by Associated Universities Inc.,
under cooperative agreement with the National Science Foundation. We
thank Dr. Kothes for useful discussions, and thank the SGPS team for
the data.

\end{document}